\documentclass[aip,reprint,floatfix]{revtex4-2}

\usepackage{graphicx}
\usepackage{amsmath,amssymb,bm}
\usepackage{siunitx}
\usepackage{hyperref}
\usepackage{microtype}
\usepackage{array}
\usepackage{placeins} 
\usepackage{float}    
\usepackage{needspace} 
\usepackage[colorinlistoftodos]{todonotes}

\graphicspath{{figs/}}

\newcommand{\dd}{\mathrm{d}}

\begin{document}

\title{An axion framework for Particle-in-Cell codes with Monte-Carlo sampling: emission, absorption, and detailed balance in plasmas}

\author{Miles Radford}
\thanks{These authors contributed equally to this work. Corresponding author: miles.radford@physics.ox.ac.uk}
\affiliation{Department of Physics, University of Oxford, Parks Road, Oxford OX1 3PU, United Kingdom}

\author{Ahmed Alsulami}
\thanks{These authors contributed equally to this work. Corresponding author: miles.radford@physics.ox.ac.uk}
\affiliation{Department of Physics, University of Oxford, Parks Road, Oxford OX1 3PU, United Kingdom}

\author{Bertrand Martinez}
\affiliation{CEA, DAM, DIF, F-91297 Arpajon, France}
\affiliation{Université Paris-Saclay, CEA, Laboratoire Matière en Conditions Extrêmes, F-91680 Bruyères-le-Châtel, France}

\author{Pablo Bilbao}
\affiliation{Department of Physics, University of Oxford, Parks Road, Oxford OX1 3PU, United Kingdom}

\author{Thomas Grismayer}
\affiliation{GoLP/Instituto de Plasmas e Fus\~ao Nuclear, Instituto Superior T\'ecnico, Universidade de Lisboa, Lisboa, Portugal}

\author{Luis O. Silva}
\affiliation{GoLP/Instituto de Plasmas e Fus\~ao Nuclear, Instituto Superior T\'ecnico, Universidade de Lisboa, Lisboa, Portugal}

\author{Robert Bingham}
\affiliation{Department of Physics, University of Strathclyde, Glasgow G4 0NG, United Kingdom}
\affiliation{Rutherford Appleton Laboratory, STFC, Didcot OX11 0QX, United Kingdom}

\author{Gianluca Gregori}
\affiliation{Department of Physics, University of Oxford, Parks Road, Oxford OX1 3PU, United Kingdom}

\date{\today}

\begin{abstract}
We present an extension of the \emph{OSIRIS} particle-in-cell (PIC) code that introduces an axion macroparticle species and three axion-production channels commonly used in thermal-plasma axion phenomenology: screened Primakoff conversion ($\gamma + Z \leftrightarrow a + Z$), Compton-like photoproduction on electrons in a blackbody photon bath ($\gamma + e \rightarrow a + e$), and thermal axion bremsstrahlung from electron--ion and electron--electron scattering ($e + Z \rightarrow e + Z + a$ and $e + e \rightarrow e + e + a$).
The package is integrated into the existing OSIRIS quantum-electrodynamics (QED) Monte Carlo infrastructure and provides Poisson macro-event sampling with unbiased weight rescaling for variance control. Optional modules implement conservative cell-local energy and momentum feedback and temperature-field evolution, and each channel includes an inverse absorption operator constructed to satisfy detailed balance with a thermal bath. We benchmark forward spectral emissivities for uniform plasmas at $T_e=\SI{1.3}{keV}$, $\SI{3}{keV}$, and $\SI{5}{keV}$ against screened analytic results based on Raffelt-style calculations, finding percent-level agreement in integrated power for all channels and good reproduction of spectral peak positions. In addition, homogeneous relaxation tests with forward and inverse operators enabled show that, for all three implemented channels, the axion population and total axion energy evolve toward stable steady-state values, providing an initial validation of detailed-balance recovery in the inverse-process implementation. These results establish a foundation for kinetic simulations of axion production, absorption, and transport in high-energy-density plasmas, while more extensive validation of feedback physics and fully dynamic multidimensional coupled scenarios remains future work.
\end{abstract}

\maketitle

\section{Introduction}
Axions were originally proposed as a dynamical solution to the strong-CP problem in quantum chromodynamics (QCD) via the Peccei--Quinn (PQ) mechanism, in which a new approximate global $\mathrm{U}(1)_{\mathrm{PQ}}$ symmetry is spontaneously broken and the associated pseudo-Nambu--Goldstone boson relaxes the effective $\theta$ parameter towards zero.\cite{peccei_quinn_1977_prl,weinberg_1978_prl,wilczek_1978_prl}
Within QCD, the resulting ``QCD axion'' acquires a mass from non-perturbative effects, with its mass and couplings controlled by a single PQ scale $f_a$.\cite{dicortona_2016_jhep,pdg_2024_rpp}
Well-studied ultraviolet axion models include hadronic/KSVZ-type models, in which the axion couples to new heavy quarks,\cite{kim_1979_prl,shifman_vainshtein_zakharov_1980_npb}
and DFSZ-type models, in which the axion couples directly to Standard-Model fermions through an extended Higgs sector.\cite{dine_fischler_srednicki_1981_plb,zhitnitsky_1980_sjn}
More generally, ``axion-like particles'' (ALPs) are pseudoscalars with similar low-energy couplings but without the strict QCD mass--coupling relation.\cite{pdg_2024_rpp,graham_2015_arnps,irastorza_redondo_2018_ppnp}

Axion phenomenology at keV-to-MeV energies is primarily governed by two effective interactions: the axion--photon coupling $g_{a\gamma\gamma}$ and the axion--electron coupling $g_{ae}$.\cite{pdg_2024_rpp}
In astrophysical and laboratory plasmas, these couplings lead to axion production and absorption processes that can be treated, at tree level, within kinetic theory.
For $g_{a\gamma\gamma}$, the key thermal-plasma channel is the Primakoff process, in which a real photon converts into an axion in the Coulomb field of charged particles (and the inverse process).\cite{Raffelt1986}
For $g_{ae}$, Compton-like photoproduction and Bremsstrahlung channels contribute, with electron--electron Bremsstrahlung becoming relevant in sufficiently hot and/or weakly coupled conditions.\cite{Raffelt1986}
A central subtlety in all three channels is plasma screening: the long-range Coulomb interaction must be regulated by an in-medium screening scale, which can substantially reduce rates relative to naive unscreened estimates.\cite{Raffelt1986}
These same processes underpin classic stellar energy-loss arguments and motivate helioscope and laboratory searches.

Experimental searches for axions and ALPs span a wide parameter space and exploit axion conversion to photons in external fields as well as direct absorption in detectors.\cite{graham_2015_arnps,irastorza_redondo_2018_ppnp,sikivie_2021_rmp}
In the dark-matter context, microwave-cavity haloscopes such as ADMX have demonstrated sensitivity to well-motivated QCD-axion parameter space in the $\mu\mathrm{eV}$ mass range.\cite{du_2018_admx}
In the solar-axion context, helioscopes use strong laboratory magnets to reconvert solar axions into x-ray photons, as exemplified by CAST and next-generation concepts such as IAXO.\cite{cast_2017_nphys,armengaud_2014_jinst_iaxo} In addition, recent laboratory searches have used x-ray free-electron-laser platforms in light-shining-through-wall experiments, exploiting Primakoff conversion and regeneration to search for axions and ALPs \cite{Halliday2025, heaton2026probing}. At the same time, high-energy-density (HED) plasma facilities provide an increasingly rich experimental environment for exploring weakly coupled new physics through controlled, transient plasmas with strong gradients in density and temperature.

Particle-in-cell (PIC) simulations are the workhorse computational tool for kinetic, multi-species plasma dynamics in HED contexts, and modern PIC codes now couple classical plasma evolution to stochastic/quantum processes via Monte Carlo operators.\cite{ridgers_2014_jcp,elkina_2011_prstab,fedeli_2022_njp_picsarqed}
OSIRIS is a mature, widely used, fully relativistic PIC code with extensive physics modules and high-performance parallel implementations.\cite{fonseca_2002_osiris}
OSIRIS also includes a Quantum-Electrodynamics (QED) Monte Carlo (MC) framework for strong-field photon emission and pair-production processes that feed back on the plasma dynamics.\cite{grismayer_2016_pop_absorption,ridgers_2014_jcp}
Recently, field-based ``axion electrodynamics'' approaches have been introduced into PIC codes by solving an axion-field equation coupled to modified Maxwell equations.\cite{an_2024_mre_axioned}
The work reported here is complementary: rather than evolving an axion field, we introduce an axion macroparticle species and implement thermal-bath axion production and absorption processes (Primakoff, Compton-like, and bremsstrahlung) as local Monte Carlo operators integrated into the OSIRIS QED--MC package.
This design targets self-consistent studies of axion emission, absorption, transport, and (optionally) plasma backreaction in spatially inhomogeneous and time-dependent plasmas, leveraging the existing QED--PIC infrastructure for weighted macroparticles and parallel diagnostics.\cite{fonseca_2002_osiris,ridgers_2014_jcp} Recent theoretical work has also shown that axion--photon coupling can drive parametric instabilities in intense electromagnetic fields, providing further motivation for the development of simulation tools capable of studying axion-related plasma phenomena in experimentally relevant settings.\cite{Beyer2023}

In this first manuscript we focus on the implementation details, on forward-process benchmarking against screened analytic calculations in the regime $T_e=\SI{1.3}{keV}$ to $\SI{5}{keV}$, and on an initial validation of the inverse operators through homogeneous equilibrium-recovery tests. In particular, we show that the coupled forward and inverse channels drive the axion population and total axion energy toward stable steady-state values for all three implemented processes. More advanced validation of conservative energy--momentum deposition, temperature evolution, and fully dynamic multidimensional coupled plasma scenarios is left for future work.

Our contribution is a unified axion particle module that (i) introduces axions as an explicit macroparticle species for transport and diagnostics, (ii) samples thermal-bath emissivities as stochastic macroparticle creation events with unbiased variance control compatible with arbitrarily-weighted PIC macroparticles, (iii) implements Primakoff conversion as a bidirectional macroparticle conversion process with explicit depletion of the parent photon/axion weight, (iv) provides inverse absorption operators for the thermal-bath channels by enforcing detailed balance against a local equilibrium spectrum, and (v) optionally couples axion emission/absorption back to the kinetic plasma through conservative cell-local energy and momentum feedback and a temperature-evolution function.

\section{Implementation methodology}
\begin{table*}[t]
\caption{Axion processes implemented in the OSIRIS QED--MC module extension in this work. ``Thermal bath'' indicates that the forward rate model assumes a local equilibrium photon (and, where needed, ion) background and does not require explicit photons/ions as emitters, while ``explicit'' indicates that the operator acts on an existing macroparticle population. All channels support optional unbiased variance control through Poisson-mean capping with weight rescaling (Sec.~\ref{sec:mc_common}), and optional conservative plasma feedback (Sec.~\ref{sec:feedback}).}
\label{tab:processes}
\centering
\newcommand{\Tcell}[2]{\parbox[t]{#1}{\raggedright\small #2}}
\begin{tabular}{@{}l l l l l@{}}
\hline\hline
\Tcell{2.3cm}{\textbf{Process}} &
\Tcell{2.2cm}{\textbf{Coupling}} &
\Tcell{3.3cm}{\textbf{Forward operator}} &
\Tcell{3.3cm}{\textbf{Inverse operator}} &
\Tcell{4.0cm}{\textbf{Notes}} \\
\hline
\Tcell{2.3cm}{Primakoff $\gamma + Z \leftrightarrow a + Z$} &
\Tcell{2.2cm}{$g_{a\gamma\gamma}$} &
\Tcell{3.3cm}{Explicit conversion acting on photon macroparticles} &
\Tcell{3.3cm}{Explicit conversion acting on axion macroparticles} &
\Tcell{4.0cm}{Debye screened Coulomb field conversion according to local cross section $\sigma$; parent macroparticle weight is depleted by converted weight; product inherits parent momentum.} \\

\Tcell{2.3cm}{Compton-like $\gamma + e \rightarrow a + e$} &
\Tcell{2.2cm}{$g_{ae}$} &
\Tcell{3.3cm}{Thermal-bath emission acting on electron macroparticles} &
\Tcell{3.3cm}{Detailed-balance absorption acting on axion macroparticles} &
\Tcell{4.0cm}{Photon bath treated as local blackbody at $T_e$; isotropic emission in bath approximation; spectrum sampled from blackbody-weighted kernel.} \\

\Tcell{2.3cm}{Bremsstrahlung $e + Z \rightarrow e + Z + a$} &
\Tcell{2.2cm}{$g_{ae}$} &
\Tcell{3.3cm}{Thermal-bath emission acting on electron macroparticles} &
\Tcell{3.3cm}{Detailed-balance absorption acting on axion macroparticles} &
\Tcell{4.0cm}{Debye screened emissivity ; optional electron--electron term $e+e\rightarrow e+e+a$.} \\
\hline\hline
\end{tabular}
\end{table*}

\subsection{Overview of OSIRIS integration}

Our axion extension follows the same architectural principle as the OSIRIS QED module : axions are introduced as a separate QED macroparticle species that can be advanced, merged, deleted, and diagnosed within the same parallel particle framework already used for other OSIRIS QED species \cite{grismayer_2016_pop_absorption,Grismayer2017SeededCascade,DelGaudio2020Compton,Amaro2026BetheHeitler,Grismayer2021VacuumPolarization,Zhang2025QuantumVacuum3D}.

\subsection{Axion species representation and kinematic assumptions}

Axions are implemented as neutral macroparticles either massless or massive and thus propagate either at $c$ or sub-luminal velocity. We present here results for the massless case and leave validation of the massive axion case for future work. For the benchmark plasmas considered here, with characteristic particle energies in the keV range, treating the axion as effectively massless is an appropriate approximation for QCD axions, whose masses are expected to be many orders of magnitude below these energy scales \cite{dicortona_2016_jhep,DiLuzio2020}. More generally, finite-mass corrections enter through $E_a=\sqrt{p_a^2+m_a^2}$ and are suppressed as $\mathcal{O}(m_a^2/E_a^2)$, so the massless limit remains accurate for light ALP scenarios with $m_a \ll \mathrm{keV}$, whereas heavier ALPs with $m_a \sim 10^2\,\mathrm{eV}$--$\mathrm{keV}$ would require a finite-mass extension of the present framework.\cite{graham_2015_arnps,irastorza_redondo_2018_ppnp,pdg_2024_rpp} In OSIRIS-normalized momentum units ($\bm{p}$ in $m_e c$), the axion energy is $E/(m_e c^2)=|\bm{p}|$ and the propagation velocity is
\begin{equation}
\bm{v} = \frac{\bm{p}}{|\bm{p}|}.
\end{equation}
The massless assumption simplifies inverse-process construction and energy--momentum bookkeeping; extensions to finite-mass kinematics (including modified emission thresholds and absorption coefficients) are implemented but validation is deferred for future work.

\subsection{Couplings and implemented channels}
\begin{figure*}[t]
\centering
\includegraphics[width=0.92\textwidth]{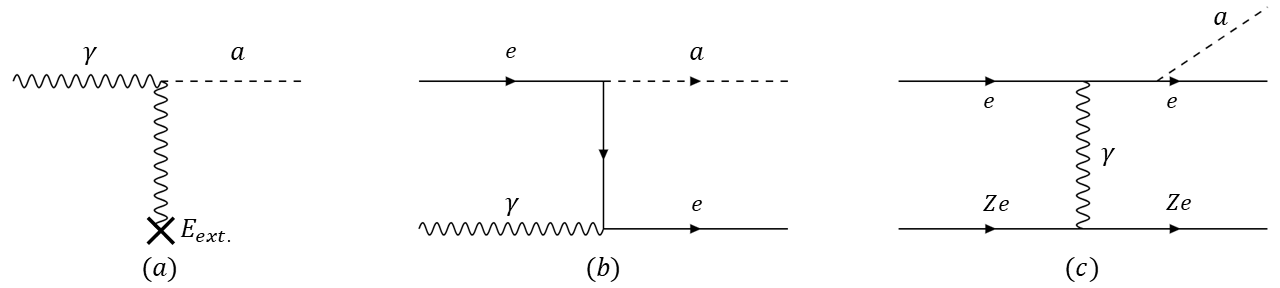}
\caption{Axion production channels implemented. (a) Primakoff conversion $\gamma + Z \rightarrow a + Z$ in an external (screened) Coulomb field. (b) Compton-like photoproduction $\gamma + e \rightarrow a + e$ in a thermal-photon bath approximation. (c) Bremsstrahlung in electron--ion scattering $e + Z \rightarrow e + Z + a$ (and analogously electron--electron scattering).}
\label{fig:channels}
\end{figure*}

We treat axions/ALPs as pseudoscalars coupled to photons and electrons through the low-energy effective interactions
\begin{equation}
\mathcal{L}_{\rm int}
= -\frac{1}{4} g_{a\gamma\gamma}\, a\, F_{\mu\nu}\tilde{F}^{\mu\nu}
+ i g_{ae}\, a\, \bar{\psi}_e \gamma_5 \psi_e,\label{eq:Lint}
\end{equation}
so that Primakoff conversion is controlled by $g_{a\gamma\gamma}$ and Compton-like and bremsstrahlung channels are controlled by $g_{ae}$.\cite{Raffelt1986,pdg_2024_rpp}
The implemented processes are shown schematically in Fig.~\ref{fig:channels} and summarised in Table~\ref{tab:processes}.

\subsection{Common Monte Carlo framework}\label{sec:mc_common}
All axion source/sink operators are formulated as local stochastic processes acting over a PIC timestep $\Delta t$ within each cell.

\subsubsection{Poisson macro-events for weighted emission}
A central requirement of the implementation is compatibility with weighted PIC macroparticles. In OSIRIS, one computational emitting particle does not represent a single physical emitter, but rather a bundle of $w_e$ physical emitters, where $w_e$ is the emitting-macroparticle weight. Let $\Gamma$ denote the physical axion-production rate per physical emitter for the process under consideration, evaluated from the local plasma state in the cell occupied by that macroparticle. Over one physical timestep $\Delta t_{\mathrm{phys}}$, the expected number of \emph{physical} axions produced by that emitting macroparticle is therefore
\begin{equation}
\langle N_a^{\mathrm{phys}} \rangle = w_e \langle \Gamma \rangle \Delta t_{\mathrm{phys}} .
\end{equation}

Creating one simulated axion for every physical axion would in general be computationally prohibitive, so emitted axions are also represented as weighted macroparticles. We denote by $w_a$ the weight assigned to a newly created axion macroparticle, i.e.\ the number of physical axions represented by one such computational particle. It is then natural to define the expected number of simulated axion-creation events associated with the emitting macroparticle during one timestep as
\begin{equation}
\lambda_{\mathrm{macro}} = \frac{w_e \Gamma \Delta t_{\mathrm{phys}}}{w_a}.
\label{eq:lambda_macro}
\end{equation}
Here $\lambda_{\mathrm{macro}}$ is the mean number of \emph{axion macroparticles} that should be created, on average, during that timestep.

The actual number of created axion macroparticles must be an integer and is sampled stochastically. We denote by $K$ the random variable giving the number of axion macroparticle creation events (``macro-events'') for the emitting macroparticle in a single timestep, and by $k=0,1,2,\ldots$ a particular realised value of $K$. In the default uncapped case, $K$ is sampled from a Poisson distribution with mean $\lambda_{\mathrm{macro}}$. More generally, to allow for the variance-control procedure described in Sec.~II.D.2, we write the sampled mean as $\lambda_{\mathrm{eff}}$, where $\lambda_{\mathrm{eff}}=\lambda_{\mathrm{macro}}$ unless Poisson-mean capping is enabled. The macro-event count is then sampled from
\begin{equation}
\Pr(K=k)=\frac{\lambda_{\mathrm{eff}}^{\,k} e^{-\lambda_{\mathrm{eff}}}}{k!},
\qquad k=0,1,2,\ldots
\label{eq:poisson_macro}
\end{equation}
which implies
\begin{equation}
\mathbb{E}[K]=\lambda_{\mathrm{eff}}, \qquad \mathrm{Var}(K)=\lambda_{\mathrm{eff}}.
\end{equation}

Operationally, one sampled macro-event corresponds to the creation of one axion macroparticle. Thus, if the draw yields $K=k$, the code creates $k$ axion macroparticles in that timestep for that emitter. In the uncapped case, each created axion macroparticle carries weight $w_a$, such that the expected number of represented physical axions is
\begin{equation}
\mathbb{E}[K]\, w_a
= \lambda_{\mathrm{macro}} w_a
= w_e \Gamma \Delta t_{\mathrm{phys}},
\end{equation}
as required by the underlying physical emission rate. This guarantees that the stochastic macro-event formulation preserves the correct mean physical yield while remaining compatible with coarse-grained weighted PIC particles.

OSIRIS operates with all quantities normalized to a reference frequency $\omega_{0}$, typically either a plasma frequency or laser frequency is chosen. Therefore the physical time associated with the leapfrog time is
\begin{equation}
\Delta t_{\mathrm{phys}} = \frac{\Delta t}{\omega_{p0}},
\label{eq:dt_phys}
\end{equation}
so that rate models expressed in SI or cgs units can be sampled consistently within the normalized PIC time advance.

\subsubsection{Unbiased variance control via Poisson-mean capping}

The expected macro-event count $\lambda_{\mathrm{macro}}$ defined in Eq.~(\ref{eq:lambda_macro}) can vary strongly across the simulation domain. If the chosen axion macroparticle weight $w_a$ is too small, then $\lambda_{\mathrm{macro}}$ may become large for a small number of high-weight emitters, leading to many Monte Carlo draws and an excessive number of created axion macroparticles in a single timestep. Conversely, choosing $w_a$ too large reduces particle-count cost but can under-resolve spectra and transport. To manage this trade-off, we implement an unbiased variance-control procedure based on capping the Poisson mean and compensating with a weight rescaling.

Specifically, instead of sampling the macro-event count with mean $\lambda_{\mathrm{macro}}$, we define an effective sampling mean
\begin{equation}
\lambda_{\mathrm{eff}} = \min(\lambda_{\mathrm{macro}}, \lambda_{\mathrm{cap}}),
\label{eq:lambda_eff}
\end{equation}
where $\lambda_{\mathrm{cap}}$ is a user-configurable upper bound. The macro-event count is then sampled exactly as in Sec.~II.D.1. 

To preserve the correct expected number of represented physical axions, the weight of each newly created axion macroparticle is increased whenever capping is active. Denoting the rescaled emitted-particle weight by $w_a^{\prime}$, we choose
\begin{equation}
w_a^{\prime} = w_a \frac{\lambda_{\mathrm{macro}}}{\lambda_{\mathrm{eff}}}.
\label{eq:wa_rescaled}
\end{equation}
This ensures that the expected physical axion yield remains unchanged:
\begin{equation}
\mathbb{E}[K]\, w_a^{\prime}
= \lambda_{\mathrm{eff}}\, w_a \frac{\lambda_{\mathrm{macro}}}{\lambda_{\mathrm{eff}}}
= \lambda_{\mathrm{macro}} w_a
= w_e \Gamma \Delta t_{\mathrm{phys}}.
\end{equation}
Thus, the cap-and-rescale procedure leaves the mean number of emitted physical axions unchanged while reducing the number of explicit Monte Carlo creation events that must be sampled.

The same argument applies to carried energy: if each created axion macroparticle is assigned an energy sampled from the same underlying emission kernel, then capping changes only the coarse-graining of the emitted ensemble, not the expected total emitted energy. The method may be viewed as importance sampling in weight space: in regions where the physical mean emission is large, the algorithm represents the same expected yield with fewer, heavier macroparticles. The broader ideas behind this construction are related to established Monte Carlo variance-reduction, weighted-particle, and population-control methods \cite{Haghighat2003,Booth2012,Legrady2020}, while the specific Poisson-mean capping and compensating weight-rescaling strategy used here is, to our knowledge, a novel adaptation within a PIC--MC framework for weighted axion emission.

We validate that this weighting and variance-control procedure is unbiased by repeating the bremsstrahlung benchmark while varying $w_a$ over several orders of magnitude. The reconstructed spectral emissivity is invariant within expected Monte Carlo noise (Appendix A), confirming that the procedure changes sampling efficiency without altering the underlying physics.

\subsection{Local plasma state, composition, and screening}

The implemented rate models assume a classical, non-degenerate Maxwellian plasma in which axion emission and absorption can be parameterised locally by the electron temperature $T_e$ and the local particle densities. For benchmarking and for astrophysical interpretability, we follow Raffelt-style notation and use the hydrogen mass fraction $X_{\mathrm H}$ together with the deposited charge density to infer the local electron density $n_e$ and mass density $\rho$, with options to assume quasi-neutrality and to bind densities to a chosen ion species.\cite{Raffelt1986}

Here $\rho$ denotes the \emph{total ion mass density} of the plasma material in the cell, not the electron mass density. In a multi-species plasma it may be written as
\begin{equation}
\rho = \sum_i n_i A_i m_u,
\label{eq:rho_def}
\end{equation}
where $n_i$ is the ion number density of species $i$, $A_i$ is its atomic mass number, and $m_u$ is the atomic mass unit. Under quasi-neutral conditions, the electron density satisfies
\begin{equation}
n_e \simeq \sum_i Z_i n_i,
\label{eq:qn_def}
\end{equation}
where $Z_i$ is the charge state of ion species $i$. Thus, $n_e$ is the free-electron number density, whereas $\rho$ is the total mass density entering the Raffelt emissivity fits used for the Compton-like and bremsstrahlung channels.

For Primakoff conversion, the screened Coulomb interaction is regulated by a Debye screening scale $\kappa$.\cite{Raffelt1986} The implementation supports both (i) a user-specified fixed $\kappa$ for controlled benchmarks and (ii) a Debye-model $\kappa$ computed from local deposited densities and $T_e$. A convenient natural-units form is
\begin{equation}
\kappa^2 = 4\pi \alpha_{\mathrm{EM}} \frac{n_{\mathrm{eff}}}{T_e},
\qquad
n_{\mathrm{eff}} = n_e + \sum_i Z_i^2 n_i,
\label{eq:kappa_debye}
\end{equation}
where $\alpha_{\mathrm{EM}} \simeq 1/137$ is the electromagnetic fine-structure constant. The quantity $n_{\mathrm{eff}}$ is the effective charge-weighted density entering the Debye screening model. Temperature may be treated in three modes: (i) globally fixed at the benchmark value (used in this manuscript), (ii) prescribed as a spatial field, or (iii) self-consistently evolved using the conservative feedback operator described in Sec.~II.J. Only mode (i) is benchmarked here.

\subsection{Primakoff conversion operator}

Primakoff conversion converts photons to axions (and vice versa) in Coulomb fields. In a plasma, screening regulates the small-angle divergence and sets the scale for the total cross section. For a photon of energy $\omega$ (with natural units where $\hbar = c = 1$) and screening scale $\kappa$, we implement the screened total cross section in the standard single-species form
\begin{equation}
\sigma_{\gamma \rightarrow a}(\omega)
= \frac{\alpha_{\mathrm{EM}} g_{a\gamma\gamma}^2 Z^2}{8}
\,F(s)\,(\hbar c)^2,
\qquad
s = \frac{4\omega^2}{\kappa^2},
\label{eq:prim_xs}
\end{equation}
with
\begin{equation}
F(s)=\left(1+\frac{1}{s}\right)\ln(1+s)-1.
\label{eq:prim_F}
\end{equation}
Here $Z$ denotes the charge state of the effective ion species entering the single-species Primakoff cross section. In the more general multi-species notation introduced above, this single-$Z$ form should be understood as a shorthand for the corresponding local ion composition, represented in the code through deposited densities and the chosen composition model.

A local target density $n_t$ is then used to convert the cross section into a per-photon conversion rate,
\begin{equation}
\Gamma_{\gamma \rightarrow a}(\omega) = n_t \,\sigma_{\gamma \rightarrow a}(\omega)\, c.
\label{eq:prim_rate}
\end{equation}
Here $n_t$ is the local number density of scattering centres responsible for Primakoff conversion. For a single ion species this is just the ion density, $n_t=n_i$. In a multi-species plasma, the factor $n_t Z^2$ is replaced by the corresponding effective charge-weighted sum over ion species.

Over one timestep, the conversion probability is sampled using the exponential survival form
\begin{equation}
P_{\gamma\rightarrow a}=1-\exp[-\Gamma_{\gamma\rightarrow a}(\omega)\Delta t_{\mathrm{phys}}].
\label{eq:prim_prob}
\end{equation}

\subsubsection{Macroparticle conversion, weight depletion, and inverse conversion}
Primakoff conversion acts on explicit photon (and axion) macroparticles.
For a parent photon macroparticle of weight $w_\gamma$, the expected number of axion macroparticles created in one step is
\begin{equation}
\lambda_P = \frac{w_\gamma\,P_{\gamma\rightarrow a}}{w_a},
\end{equation}
with Poisson sampling and (optional) cap-and-rescale variance control as in Sec.~\ref{sec:mc_common}.
Each conversion event creates a product axion macroparticle that inherits the parent momentum ($\bm{p}_a\leftarrow\bm{p}_\gamma$), ensuring energy-momentum conservation at the macroparticle level.
The parent photon macroparticle weight is depleted by the converted physical-particle weight, guaranteeing global number conservation across the conversion operator.

The inverse conversion $a\rightarrow\gamma$ is implemented analogously with the same screened kernel and the same depletion logic, so that the pair of operators constitute a bidirectional macroparticle-conversion process. The equilibrium behaviour of this coupled $\gamma \leftrightarrow a$ operator is validated in Sec.~III.F through homogeneous relaxation tests, which show convergence of both axion population and total axion energy toward steady-state values.

\subsection{Compton-like photoproduction operator (thermal bath)}

The Compton-like channel $\gamma + e \rightarrow a + e$ depends on the axion--electron coupling $g_{ae}$ and is relevant for DFSZ-like axions and generic ALPs.\cite{pdg_2024_rpp,Raffelt1986} We implement this process in a thermal-bath approximation: electrons are kinetic PIC particles, while bath photons are not explicitly simulated and are instead modelled as a local blackbody spectrum at temperature $T_e$.
Following Raffelt, the total axion energy-loss rate per unit mass can be expressed as
\begin{eqnarray}
\epsilon_C(T_e, X_{\mathrm H}) \simeq 2.67 \times 10^{-2}\,
\mathrm{erg\,g^{-1}\,s^{-1}} \times \nonumber \\
(1 + X_{\mathrm H})\,T_7^6
\left(\frac{\alpha_{ae}}{1.60\times10^{-23}}\right),
\label{eq:epsC}
\end{eqnarray}
where $T_7 \equiv T_e/(10^7\,\mathrm{K})$ and $\alpha_{ae} \equiv g_{ae}^2/(4\pi)$.

In the code, $\epsilon_C$ is converted to a volumetric emissivity through
\begin{equation}
Q_C = \rho\,\epsilon_C,
\label{eq:QC}
\end{equation}
where $\rho$ is the total plasma mass density defined in Sec.~II.E, not the electron mass density. The per-electron production rate is then estimated as
\begin{equation}
\Gamma_C \approx \frac{Q_C}{n_e \langle E \rangle},
\label{eq:GammaC}
\end{equation}
where $n_e$ is the free-electron number density and $\langle E \rangle$ is the assumed mean emitted axion energy used to convert the energy emissivity into a number-production rate. Thus, $\rho$ and $n_e$ play different roles: $\rho$ appears because Raffelt quotes the axion rate per unit mass, while $n_e$ converts the resulting volumetric emissivity into a rate per emitting free electron.

As implemented here, this operator assumes a classical, non-degenerate, effectively fully ionised plasma and therefore treats Compton-like axion production as scattering from free electrons only. Contributions from bound electrons are neglected, consistent with the thermal-plasma regime and Raffelt-style emissivity model used for the present benchmarks. Macro-event emission, energy sampling, and optional variance control then follow the common framework of Sec.~II.D.

For spectral sampling, we draw axion energies from an analytic kernel consistent with the thermal Compton approximation, which for the power spectrum scales as
\begin{equation}
\frac{dQ_C}{dE} \propto \frac{E^5}{\exp(E/T_e)-1}.
\label{eq:compton_kernel}
\end{equation}
Emission directions are sampled isotropically in this approximation.

\subsection{Bremsstrahlung operator and electron--electron contribution}
Thermal axion bremsstrahlung is implemented as a local emission operator acting on electrons, using screened analytic emissivity models for electron--ion bremsstrahlung and an optional electron--electron term.\cite{Raffelt1986}
Using Raffelt's fitted energy-loss-rate expressions as a reference, the electron--ion contribution can be written in terms of $T_7\equiv T_e/(10^7\,\mathrm{K})$ and $\rho_2\equiv\rho/(100\,\mathrm{g\,cm^{-3}})$ as in \cite{Raffelt1986}. Here, as in the Compton-like operator, $\rho$ denotes the total plasma mass density defined in Sec.~II.E.
\begin{widetext}
\begin{equation}
\begin{aligned}
\epsilon_{eZ}(\rho,T_e,X_H) \simeq {}&
\left[0.150\,\mathrm{erg\,g^{-1}\,s^{-1}}\right]\,T_7^{2.5}\,\rho_2\,(1+X_H)
\left(\frac{\alpha_{ae}}{1.60\times10^{-23}}\right)
\left[1-6.98\times10^{-2}\,\eta\right],
\end{aligned}
\label{eq:eps_eZ}
\end{equation}
\end{widetext}
where $\eta\equiv (3+X_H)\rho_2 T_7^{-2}$ captures the screening-correction fit. Here the factor $1 - 6.98 \times 10^{-2}\eta$ is the screening correction appearing in Raffelt's fitted thermal bremsstrahlung rate model, with $\eta \equiv (3 + X_{\mathrm H}) \rho_2 T_7^{-2}$ used as the fit variable. In the present implementation, this correction is therefore not introduced through an explicit differential matrix element regulated by the Debye scale $\kappa$ as in the Primakoff operator. Instead, screening enters implicitly through Raffelt's emissivity fit, in which the dependence on the underlying plasma screening physics has already been absorbed into the fitted $\eta$ dependence. Thus, $\eta$ and $\kappa$ are not used as interchangeable quantities here: $\kappa$ appears explicitly in the Primakoff cross section, whereas for bremsstrahlung the screening correction is inherited from the fitted emissivity model itself. When enabled, the electron--electron contribution is\cite{Raffelt1986}
\begin{widetext}
\begin{equation}
\begin{aligned}
\epsilon_{ee}(\rho,T_e,X_H) \simeq {}&
\left[0.150\,\mathrm{erg\,g^{-1}\,s^{-1}}\right]\,T_7^{2.5}\,\rho_2\,(1+X_H)
\frac{1+X_H}{2\sqrt{2}}
\left(\frac{\alpha_{ae}}{1.60\times10^{-23}}\right)
\left[1-0.140\,\eta\right].
\end{aligned}
\label{eq:eps_ee}
\end{equation}
\end{widetext}
The total bremsstrahlung emissivity is $\epsilon_B=\epsilon_{eZ}+\epsilon_{ee}$ when the $ee$ term is enabled.
As for Compton-like emission, $\epsilon_B$ is converted to $Q_B=\rho\epsilon_B$ and then to a per-electron production rate $\Gamma_B\approx Q_B/(n_e\langle E\rangle)$.

For energy sampling we draw the dimensionless axion energy $w \equiv E/T_e$ from a gamma distribution. In the present implementation, this Gamma-distribution sampler is tuned to reproduce the peak location and mean energy of the screened bremsstrahlung spectrum implied by the Raffelt-based emissivity model used for the total rate. It is therefore not an independent analytic model, but a numerical sampling surrogate constructed to approximate the screened bremsstrahlung spectrum efficiently over the benchmark temperature range considered here. Emission directions are sampled isotropically in the bath approximation.

\subsection{Inverse processes and detailed balance}
Beyond explicit Primakoff $a\leftrightarrow\gamma$ conversion, we implement inverse \emph{absorption} operators for the thermal-bath Compton-like and bremsstrahlung channels using detailed balance.
The guiding requirement is that, in a static homogeneous cell with fixed $T_e$ and densities, forward emission and inverse absorption drive the axion distribution towards thermal equilibrium with the thermal bath.

The forward model defines a differential number emissivity $\dd \dot{n}_a/\dd E$.
We choose a target equilibrium axion spectrum for a (nearly) massless boson with zero chemical potential, with an option to use either Bose--Einstein (BE) or Maxwell--Boltzmann (MB) statistics:
\begin{align}
\frac{\dd n_a^{\mathrm{eq}}}{\dd E} &= \frac{E^2}{2\pi^2(\hbar c)^3}\,\frac{1}{\exp(E/T_e)-1}\quad(\mathrm{BE}),\\
\frac{\dd n_a^{\mathrm{eq}}}{\dd E} &= \frac{E^2}{2\pi^2(\hbar c)^3}\,\exp(-E/T_e)\quad(\mathrm{MB}).
\end{align}
Detailed balance is enforced by defining an absorption rate
\begin{equation}
\Gamma_{\mathrm{abs}}(E)=\frac{(\dd \dot{n}_a/\dd E)}{\dd n_a^{\mathrm{eq}}/\dd E},
\label{eq:gamma_abs}
\end{equation}
so that the net source vanishes energy-bin-by-energy-bin when $n_a=n_a^{\mathrm{eq}}$.
Absorption is sampled using $P_{\mathrm{abs}}=1-\exp(-\Gamma_{\mathrm{abs}}\Delta t_{\rm phys})$ per axion macroparticle, with weight-aware deletion or weight reduction.

\subsection{Energy--momentum feedback and temperature evolution}\label{sec:feedback}
A principal motivation for embedding axion operators inside OSIRIS QED--MC is the ability to couple axion emission and absorption to the plasma evolution.
In addition to event-level conservation where applicable (e.g., photon-to-axion conversion), the code supports cell-local energy and momentum exchange buffers that accumulate $(\Delta E,\Delta\bm{P})$ transferred to or from axions during a timestep.
These buffers enable a conservative feedback mapping back onto the plasma macroparticles after MC sampling, avoiding the requirement for the consideration of recoil at the individual particle level.

\subsubsection{Affine momentum remapping in a cell}
For the thermal-bath channels, the Monte Carlo operators do not resolve microscopic recoil kinematics event-by-event. Instead, they accumulate the net energy and momentum that should be transferred between the plasma and the axion population in each cell over one timestep. For a given cell, let $\Delta E$ denote the desired total kinetic-energy change to be applied to the chosen plasma species, and let $\Delta \mathbf{P}=(\Delta P_x,\Delta P_y,\Delta P_z)$ denote the corresponding desired total momentum change, as obtained from the cell-local Monte Carlo bookkeeping of emission and absorption events.

Let the cell contain particles with momenta $\mathbf{p}_i =
(p_{x,i},p_{y,i},p_{z,i})$ and effective particle weights $w_i$,
and define the weighted moments
\begin{equation}
W_c = \sum_i w_i, \qquad
\mathbf{P} = \sum_i w_i \mathbf{p}_i, \qquad
U_2 = \sum_i w_i |\mathbf{p}_i|^2 .
\label{eq:cell_moments}
\end{equation}
The weighted mean momentum is then $\bar{\mathbf{p}}=\mathbf{P}/W_c$.
We further define the \emph{unnormalized weighted central second moment}
of the momentum distribution by
\begin{equation}
\mathrm{var}
\equiv
\sum_i w_i |\mathbf{p}_i-\bar{\mathbf{p}}|^2
=
U_2-\frac{|\mathbf{P}|^2}{W_c}.
\label{eq:var_unnormalized}
\end{equation}
Equivalently, the corresponding normalized weighted variance is
\begin{equation}
\frac{\mathrm{var}}{W_c}
=
\frac{U_2}{W_c}
-
\frac{|\mathbf{P}|^2}{W_c^2}.
\label{eq:var_normalized}
\end{equation}
In the present implementation, the feedback operator is based on a
nonrelativistic kinetic-energy model, so $\mathrm{var}$ is used in its
unnormalized form, which is proportional to the total thermal part of the
cell kinetic energy after subtraction of the bulk flow.

Rather than applying many small stochastic kicks particle-by-particle, the code remaps all momenta in the cell using an affine transformation,
\begin{equation}
\mathbf{p}_i' = \mathcal{S}\,\mathbf{p}_i + \mathbf{b},
\label{eq:affine_map}
\end{equation}
where $S$ is a scalar rescaling factor and $\mathbf{b}=(b_x,b_y,b_z)$ is a uniform momentum shift. The momentum part of the feedback is first written as a per-effective-particle translation,
\begin{equation}
\boldsymbol{\delta}=(\delta_x,\delta_y,\delta_z)
=\frac{\Delta \mathbf{P}}{W_c}.
\label{eq:delta_def}
\end{equation}
A pure translation by $\boldsymbol{\delta}$ would already change the cell kinetic energy, so the associated bulk contribution is computed explicitly as
\begin{equation}
\Delta E_{\mathrm{bulk}}
=
\frac{1}{2\mu}
\left[
2\,\boldsymbol{\delta}\!\cdot\!\mathbf{P}
+
W_c |\boldsymbol{\delta}|^2
\right],
\label{eq:dE_bulk}
\end{equation}
where $\mu$ is the particle mass in units of $m_e$. The remaining energy change that must be carried by a change in the thermal spread is then
\begin{equation}
\Delta E_{\mathrm{th}}=\Delta E-\Delta E_{\mathrm{bulk}}.
\label{eq:dE_th}
\end{equation}

The target post-remap variance is chosen as
\begin{equation}
\mathrm{var}_{\mathrm{new}}
=
\mathrm{var}+2\mu\,\Delta E_{\mathrm{th}},
\label{eq:var_new}
\end{equation}
with $\mathrm{var}_{\mathrm{new}}$ reduced to zero if necessary. If $\mathrm{var}>0$, the momentum-rescaling factor is then
\begin{equation}
S=\sqrt{\frac{\mathrm{var}_{\mathrm{new}}}{\mathrm{var}}},
\label{eq:s_def}
\end{equation}
while for an effectively cold cell the code sets $S=1$ and leaves the variance unchanged. Once $S$ is known, the offset vector is chosen as
\begin{equation}
\mathbf{b}=(1-S)\bar{\mathbf{p}}+\boldsymbol{\delta},
\label{eq:b_def}
\end{equation}
so that the new mean momentum becomes
\begin{equation}
\bar{\mathbf{p}}' = S\,\bar{\mathbf{p}}+\mathbf{b}
= \bar{\mathbf{p}}+\boldsymbol{\delta},
\end{equation}
and therefore the total cell momentum changes by exactly $\Delta \mathbf{P}=W_c\boldsymbol{\delta}$.

This construction separates the two roles of the remap: the translation $\boldsymbol{\delta}$ enforces the required momentum transfer, while the rescaling $S$ adjusts the thermal spread so that the remaining energy transfer is applied conservatively at the cell level thus accounting for both bulk and peculiar flow. This provides a low-noise way of coupling unresolved thermal axion source and sink terms back to the resolved PIC particle distribution.

\subsubsection{Temperature estimation and evolution}
When temperature evolution is enabled, the local electron temperature used by the thermal-bath operators is updated from the post-remap cell moments. In the present implementation, the temperature estimate is obtained from the central second moment of the momentum distribution after bulk-flow subtraction, using the same nonrelativistic approach as in the feedback operator. Specifically, after computing $\mathrm{var}_{\mathrm{new}}$ from Eq.~(\ref{eq:var_new}), the code estimates
\begin{equation}
T_{e,\mathrm{est}}
=
\frac{m_e c^2}{3}
\frac{\mathrm{var}_{\mathrm{new}}}{W_c},
\label{eq:Te_est}
\end{equation}
where $m_e c^2=511$~keV in physical units. Thus, $T_{e,\mathrm{est}}$ is an effective cell temperature inferred from the random momentum spread, rather than from the total kinetic energy including bulk drift.

To avoid abrupt temperature jumps from Monte Carlo noise, the stored cell temperature is not replaced instantaneously by $T_{e,\mathrm{est}}$. Instead, it is relaxed toward that value according to
\begin{equation}
T_e^{\,n+1}
=
(1-r)\,T_e^{\,n}
+
r\,T_{e,\mathrm{est}},
\label{eq:Te_relax}
\end{equation}

where $r\in[0,1]$ is a user-configurable relaxation parameter controlling the response time of the temperature update. When applied every $\Delta t_{\mathrm{upd}}$, it corresponds to an effective relaxation timescale
\[
\tau_{\mathrm{rel}} = -\frac{\Delta t_{\mathrm{upd}}}{\ln(1-r)}
\approx \frac{\Delta t_{\mathrm{upd}}}{r}\qquad (r\ll 1).
\]
Thus $r=1$ gives an instantaneous update, whereas smaller $r$ enforces a slower relaxation toward the estimated temperature. The updated temperature is then translated to user-specified floor and ceiling values before being stored for the next timestep.

In this way, the feedback loop proceeds as follows: the Monte Carlo axion operators determine the net cell-integrated energy and momentum exchange, the affine remapping applies that exchange conservatively to the particle distribution, and the local temperature used by subsequent thermal rates is updated from the remapped distribution itself. This provides a simple self-consistent pathway for axion emission and absorption to modify the local plasma state without explicitly resolving individual recoil events.

\subsection{Current limitations}

The work presented here makes several approximations whose impact should be kept in mind when interpreting the results. First, the work in this manuscript treats axions as massless. For the keV-scale thermal plasmas benchmarked here, this has negligible impact on the reported spectra and integrated emissivities provided $m_a \ll E_a$, since finite-mass corrections then remain parametrically small. However, for heavier ALPs with non-negligible kinematic corrections or threshold effects, a finite-mass prescription is required. 

Second, the Compton-like and bremsstrahlung channels are implemented in a thermal-bath approximation using local rate models and simplified emission kernels. In the homogeneous thermal benchmarks considered in this work, this does not prevent accurate recovery of the expected emissivity spectra in an averaged sense, as demonstrated by the agreement with the Raffelt results. The main limitation is therefore not the benchmark accuracy itself, but the range of plasma conditions for which the model can be expected to remain valid. In particular, non-thermal kinetic physics.

Third, the underlying screened emissivity models are for classical, weakly coupled, and non-degenerate thermal-plasmas. The overall accuracy of the forward benchmarks is therefore tied to the validity of those assumptions: in the regime of homogeneous keV thermal plasmas studied here, the agreement with the analytic reference spectra indicates that the implementation is accurate at the level of the chosen physical model, whereas strongly degenerate, strongly coupled, highly relativistic, or strongly non-thermal plasmas would require more general rate prescriptions.

Finally, although inverse operators, conservative energy--momentum feedback, affine momentum remapping, and temperature evolution are implemented, the present validation covers the forward emissivity operators and the homogeneous equilibrium-recovery behaviour of the inverse channels. The remaining plasma-feedback components should therefore be regarded as implemented capabilities whose detailed validation in self-consistent coupled dynamic plasma scenarios remains future work.

It is also useful to distinguish the present approach from PIC implementations based on axion electrodynamics, in which the axion is introduced as a coherent field coupled directly to modified Maxwell equations. Such field-based implementations have already been developed in PIC codes and are naturally suited to phase-coherent phenomena, including axion--photon mixing at the field level and instability problems in which the axion behaves as an additional dynamical wave field.\cite{an_2024_mre_axioned} By contrast, the present module treats axions as stochastic kinetic macroparticles produced, absorbed, and transported through local Monte Carlo operators. This makes it naturally suited to incoherent thermal emission, inverse processes, transport, and particle-based plasma feedback.

Ultimately, the most complete framework would combine both descriptions within a single code, allowing coherent axion-field dynamics and incoherent particle emission/absorption to be treated simultaneously. However, such a hybrid formulation must be constructed carefully to avoid double counting regions of axion phase space represented in both field and particle form. In practice, this requires a consistent prescription for transferring energy and occupancy between the coherent field sector and the kinetic particle sector, so that the combined description remains conservative and physically accurate \cite{gonoskov2015extended}.

\section{Benchmarking results}
\subsection{Benchmark procedure}
We benchmark each forward axion-production operator using a uniform, static thermal plasma with fixed electron temperature $T_e \in \{\SI{1.3}{keV},\SI{3}{keV},\SI{5}{keV}\}$.
For each channel, OSIRIS is run with only the corresponding axion operator active.
The analytic comparison curves correspond to screened Raffelt-style thermal-plasma emissivities for Primakoff, Compton-like production, and bremsstrahlung (including screening reductions and, where relevant, composition dependence).\cite{Raffelt1986}

\subsection{Absolute spectral emissivities: temperature scan}
\begin{figure}[H]
\centering
\includegraphics[width=\linewidth]{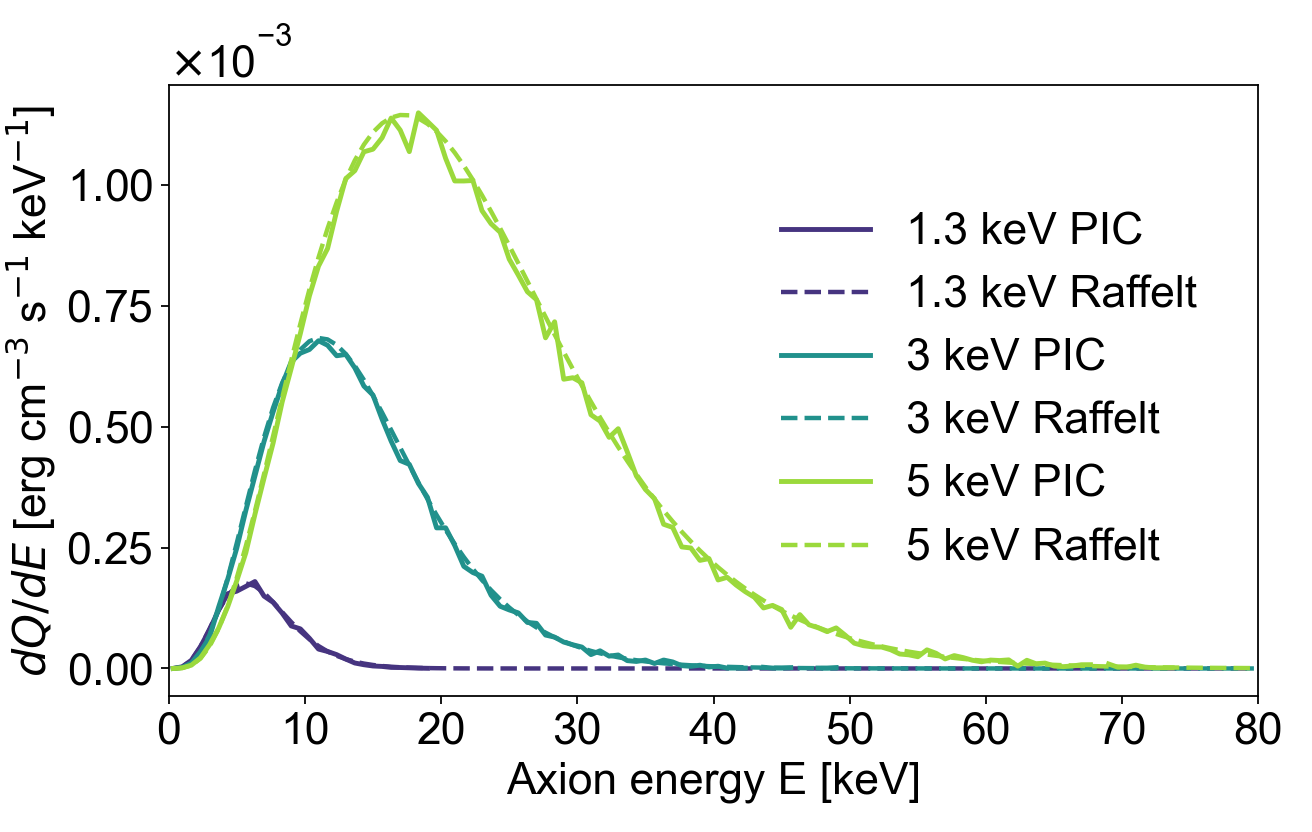}
\caption{Primakoff benchmark: volumetric spectral emissivity $\dd Q/\dd E$ for $T_e=\SI{1.3}{keV},\SI{3}{keV},\SI{5}{keV}$. Solid: OSIRIS axion module. Dashed: screened analytic benchmark (Raffelt-style).\cite{Raffelt1986}}
\label{fig:primakoff}
\end{figure}

Figure~\ref{fig:primakoff} shows the Primakoff $\dd Q/\dd E$ spectrum as a function of axion energy for the three benchmark temperatures, with OSIRIS (solid) compared against the screened analytic curve (dashed). For the Primakoff benchmarks shown in ~\ref{fig:primakoff}, the coupling is fixed at $g_{a\gamma\gamma} = 10^{-13} GeV^{-1}$. At fixed plasma conditions, the Primakoff emissivity is expected to scale quadratically with this coupling, $dQ/dE \propto g_{a\gamma\gamma}^2$.

\begin{figure}[H]
\centering
\includegraphics[width=\linewidth]{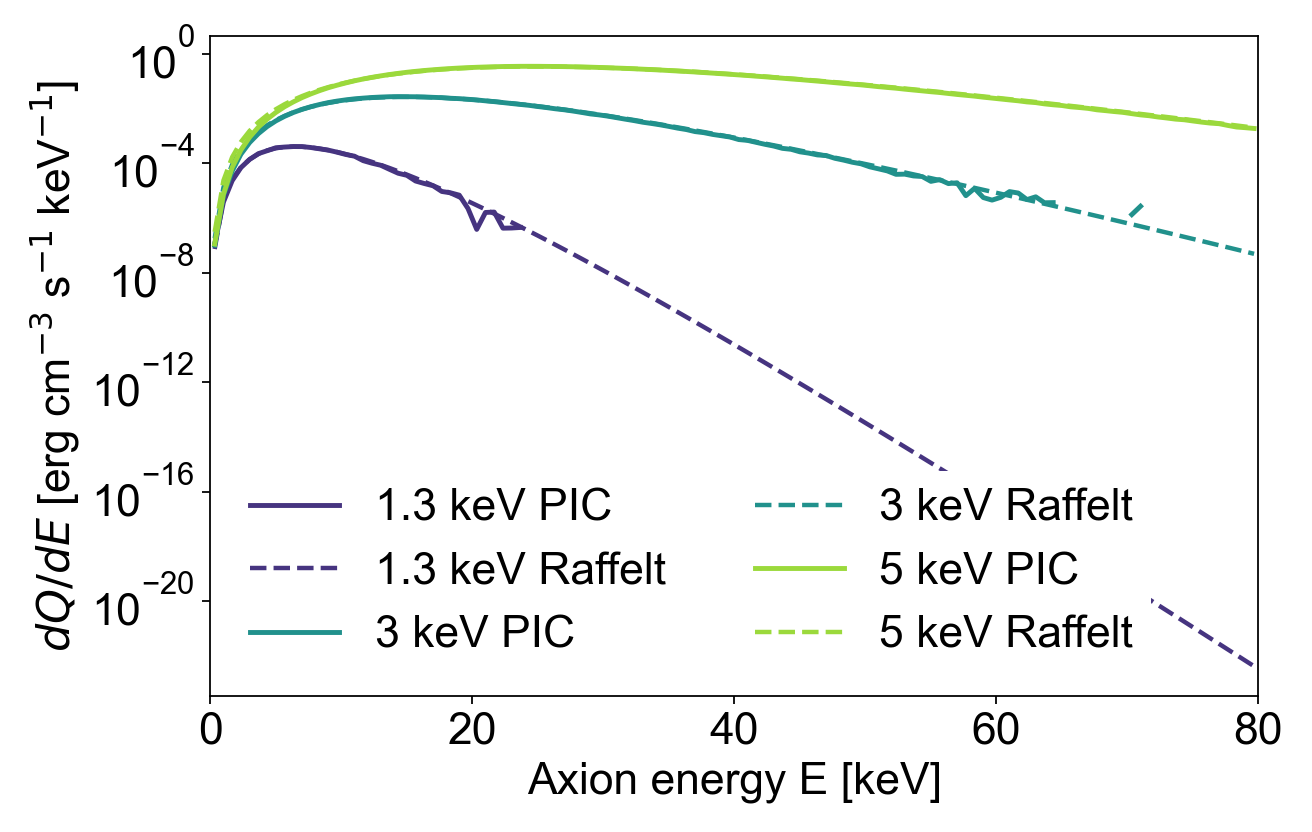}
\caption{Compton-like benchmark (logarithmic $y$-axis): volumetric spectral emissivity $\dd Q/\dd E$ for $T_e=\SI{1.3}{keV},\SI{3}{keV},\SI{5}{keV}$. The strong temperature scaling of the Compton-like rate produces a wide dynamic range; the log scale allows all three curves to be displayed. Solid: OSIRIS. Dashed: screened analytic benchmark.\cite{Raffelt1986}}
\label{fig:compton}
\end{figure}

Figure~\ref{fig:compton} shows the Compton-like photoproduction benchmark. Because the emissivity depends steeply on temperature, we use a logarithmic $y$-axis so that all three temperature curves are visible on the same panel. For the Compton benchmarks shown in ~\ref{fig:compton}, the coupling is fixed at $g_{ae}=10^{-13}$. At fixed plasma conditions, the Compton (and Bremsstrahlung) emissivity is expected to scale quadratically with this coupling, $dQ/dE \propto g_{ae}^{\,2}$, since the rate depends on $\alpha_{ae}\equiv g_{ae}^2/(4\pi)$.

\begin{figure}[H]
\centering
\includegraphics[width=\linewidth]{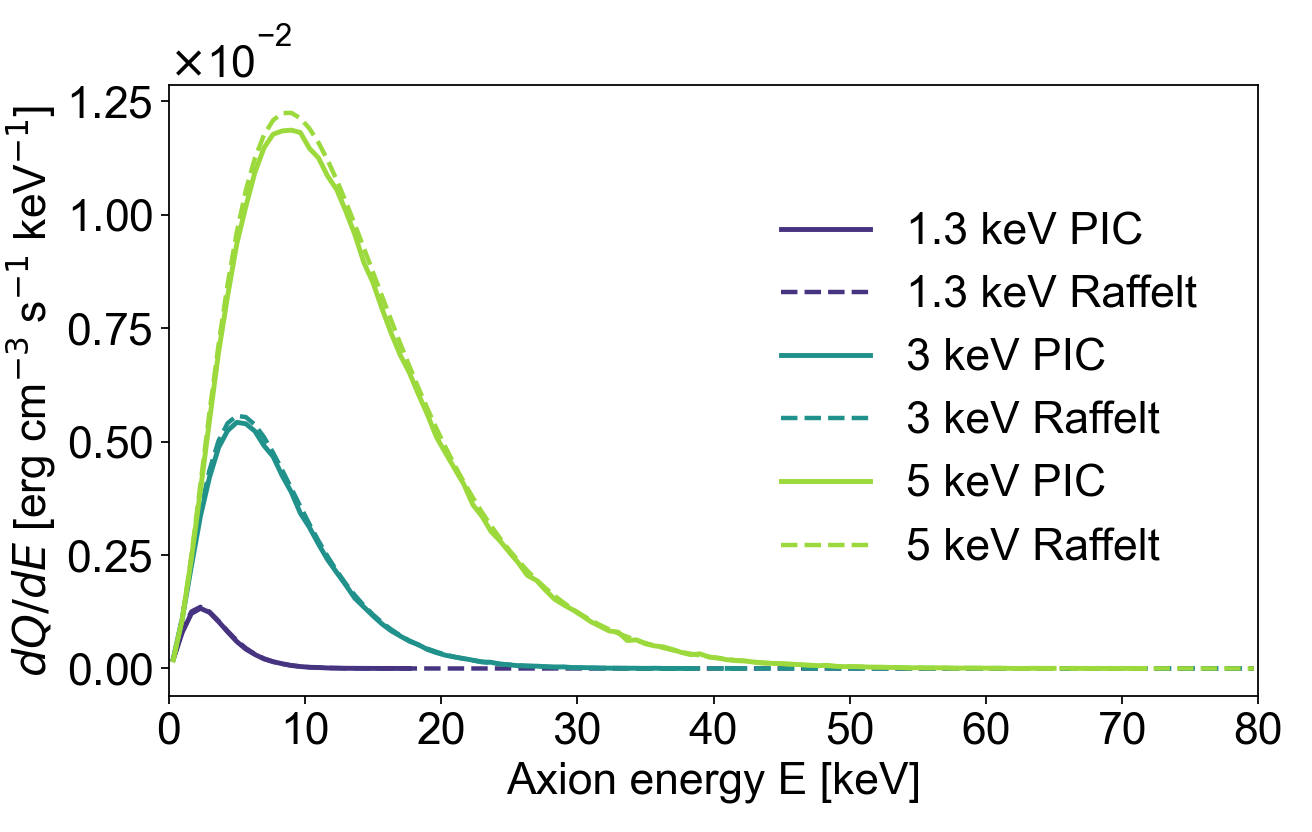}
\caption{Bremsstrahlung benchmark: volumetric spectral emissivity $\dd Q/\dd E$ for $T_e=\SI{1.3}{keV},\SI{3}{keV},\SI{5}{keV}$. Solid: OSIRIS. Dashed: screened analytic benchmark.\cite{Raffelt1986}}
\label{fig:brems}
\end{figure}

Figure~\ref{fig:brems} shows the bremsstrahlung channel benchmark.

\begin{figure}[H]
\centering
\includegraphics[width=\linewidth]{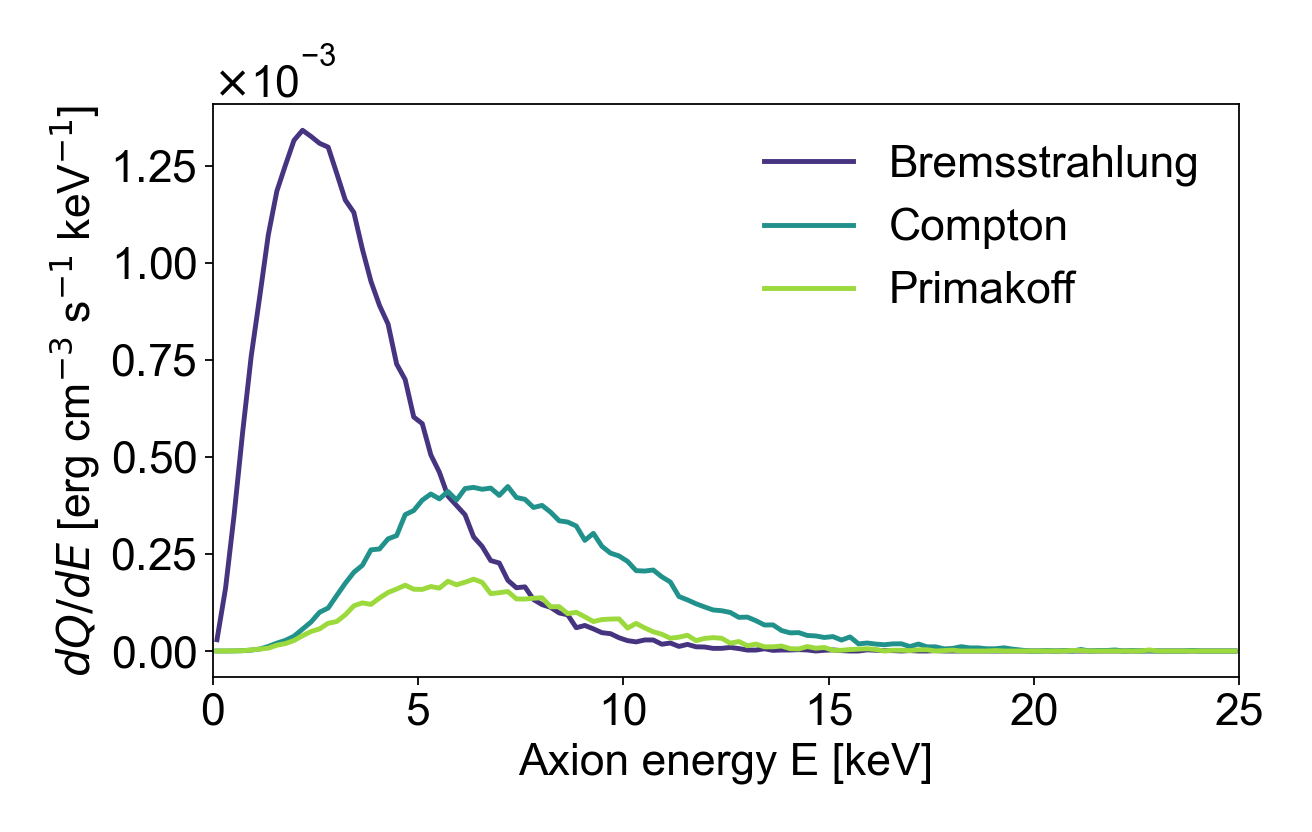}
\caption{PIC-only overlay at $T_e=\SI{1.3}{keV}$ showing the relative magnitudes and spectral locations of bremsstrahlung, Compton-like, and Primakoff emissivities in the homogeneous-box benchmark configuration.}
\label{fig:overlay}
\end{figure}

To illustrate the relative importance of the three channels at a representative low temperature, Fig.~\ref{fig:overlay} overlays the PIC-only spectra at $T_e=\SI{1.3}{keV}$.

\begin{table*}[t]
\caption{Quantitative benchmark summary for the forward-process spectral emissivity $\dd Q/\dd E$ at $T_e=\SI{1.3}{keV},\,\SI{3}{keV},\,\SI{5}{keV}$. ``PIC'' spectra are obtained from axion macroparticles produced by the corresponding OSIRIS MC operator, while ``Raffelt'' denotes the screened analytic benchmark.\cite{Raffelt1986} Peak locations and heights are defined from discretised curves. The integrated powers are defined as $Q_{\mathrm{PIC}}=\int (\dd Q/\dd E)_{\mathrm{PIC}}\,\dd E$ and $Q_{R}=\int (\dd Q/\dd E)_{R}\,\dd E$, evaluated over $E\in[0,80]$~keV, where $Q_{\mathrm{PIC}}$ and $Q_R$ are the total volumetric emissivities of the PIC and Raffelt spectra, respectively.}
\label{tab:bench_summary}
\setlength{\tabcolsep}{3pt}
\renewcommand{\arraystretch}{0.92}
\small
\begin{ruledtabular}
\begin{tabular}{l c c c c c c c c c}
Process & $T_e$ [keV] & $E_{\mathrm{pk}}^{\mathrm{PIC}}$ [keV] & $(\dd Q/\dd E)_{\mathrm{pk}}^{\mathrm{PIC}}$ &
$E_{\mathrm{pk}}^{\mathrm{R}}$ [keV] & $(\dd Q/\dd E)_{\mathrm{pk}}^{\mathrm{R}}$ &
$Q^{\mathrm{PIC}}$ & $Q^{\mathrm{R}}$ &
$Q^{\mathrm{PIC}}/Q^{\mathrm{R}}$ & rel.\ $L_2$ \\
\hline
Bremsstrahlung & \num{1.30} & \num{2.240} & \num{1.72e-3} & \num{2.240} & \num{1.72e-3} & \num{3.17e-2} & \num{3.17e-2} & \num{1.00} & \num{0.00} \\
Bremsstrahlung & \num{3.00} & \num{5.039} & \num{5.37e-3} & \num{5.039} & \num{5.50e-3} & \num{8.89e-2} & \num{8.91e-2} & \num{0.997} & \num{1.58e-2} \\
Bremsstrahlung & \num{5.00} & \num{8.679} & \num{1.11e-2} & \num{8.469} & \num{1.15e-2} & \num{2.20e-1} & \num{2.22e-1} & \num{0.992} & \num{2.40e-2} \\
Primakoff & \num{1.30} & \num{6.299} & \num{2.24e-4} & \num{6.089} & \num{2.24e-4} & \num{5.08e-3} & \num{5.08e-3} & \num{1.00} & \num{6.56e-3} \\
Primakoff & \num{3.00} & \num{11.059} & \num{6e-4} & \num{11.479} & \num{7.01e-4} & \num{1.35e-2} & \num{1.35e-2} & \num{1.00} & \num{1.62e-2} \\
Primakoff & \num{5.00} & \num{18.408} & \num{1.14e-3} & \num{16.868} & \num{1.14e-3} & \num{2.97e-2} & \num{2.99e-2} & \num{0.992} & \num{2.34e-2} \\
Compton & \num{1.30} & \num{6.369} & \num{4.27e-4} & \num{6.369} & \num{4.27e-4} & \num{3.06e-3} & \num{3.06e-3} & \num{0.999} & \num{2.02e-3} \\
Compton & \num{3.00} & \num{13.954} & \num{2.83e-2} & \num{13.954} & \num{2.83e-2} & \num{4.71e-1} & \num{4.71e-1} & \num{1.000} & \num{3.13e-4} \\
Compton & \num{5.00} & \num{21.395} & \num{3.50e-1} & \num{21.395} & \num{3.50e-1} & \num{1.01e+1} & \num{1.01e+1} & \num{1.000} & \num{3.27e-4} \\
\end{tabular}
\end{ruledtabular}
\end{table*}


\subsection{Electron--electron contribution in bremsstrahlung}
\begin{figure}[H]
\centering
\includegraphics[width=\linewidth]{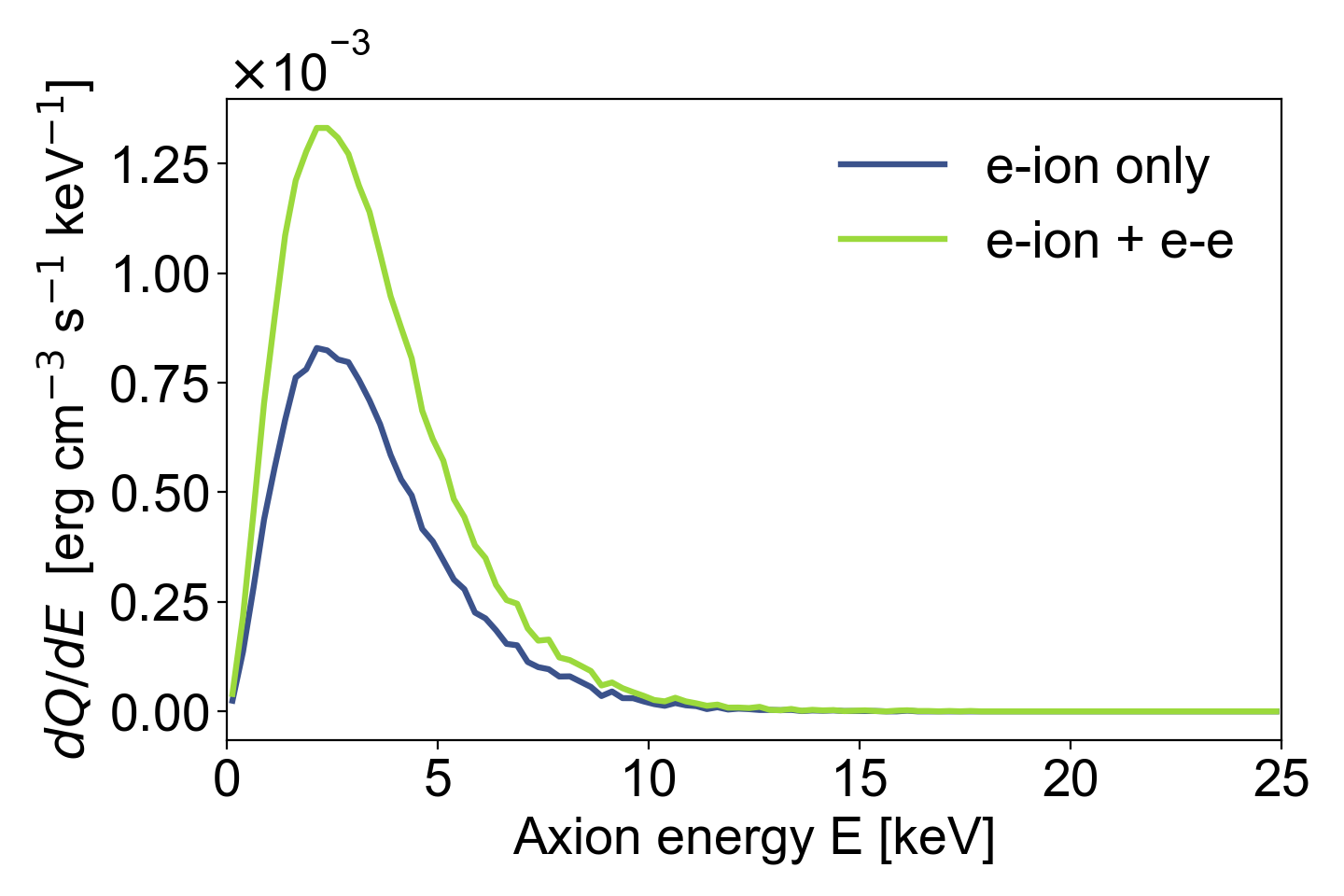}
\caption{Bremsstrahlung $\dd Q/\dd E$: electron--ion only versus electron--ion plus electron--electron. \cite{Raffelt1986}}
\label{fig:brems_ei_ee}
\end{figure}

Figure~\ref{fig:brems_ei_ee} compares bremsstrahlung spectra obtained with only the electron--ion contribution retained against spectra obtained with both the electron--ion and electron--electron terms included. In Raffelt-style thermal-plasma calculations, the $e$--$e$ channel can make a non-negligible and sometimes substantial contribution to the total axion bremsstrahlung emissivity. The purpose of Fig.~6 is therefore to show that the OSIRIS implementation accounts for this extra channel and reproduces the corresponding increase in emissivity in general agreement with the Raffelt's reference calculation.

\subsection{Quantitative agreement}
Table~\ref{tab:bench_summary} summarises peak energies, peak values, integrated emissivities, PIC-to-analytic ratios, and a relative $L_2$ residual measure.
Across this temperature scan, the total power agrees at the percent level for all channels, consistent with the design goal that the MC operators reproduce the underlying screened emissivity models in expectation.\cite{Raffelt1986}

\subsection{Equilibrium recovery with forward and inverse operators}

\begin{figure}[t]
    \centering
    \includegraphics[width=\linewidth]{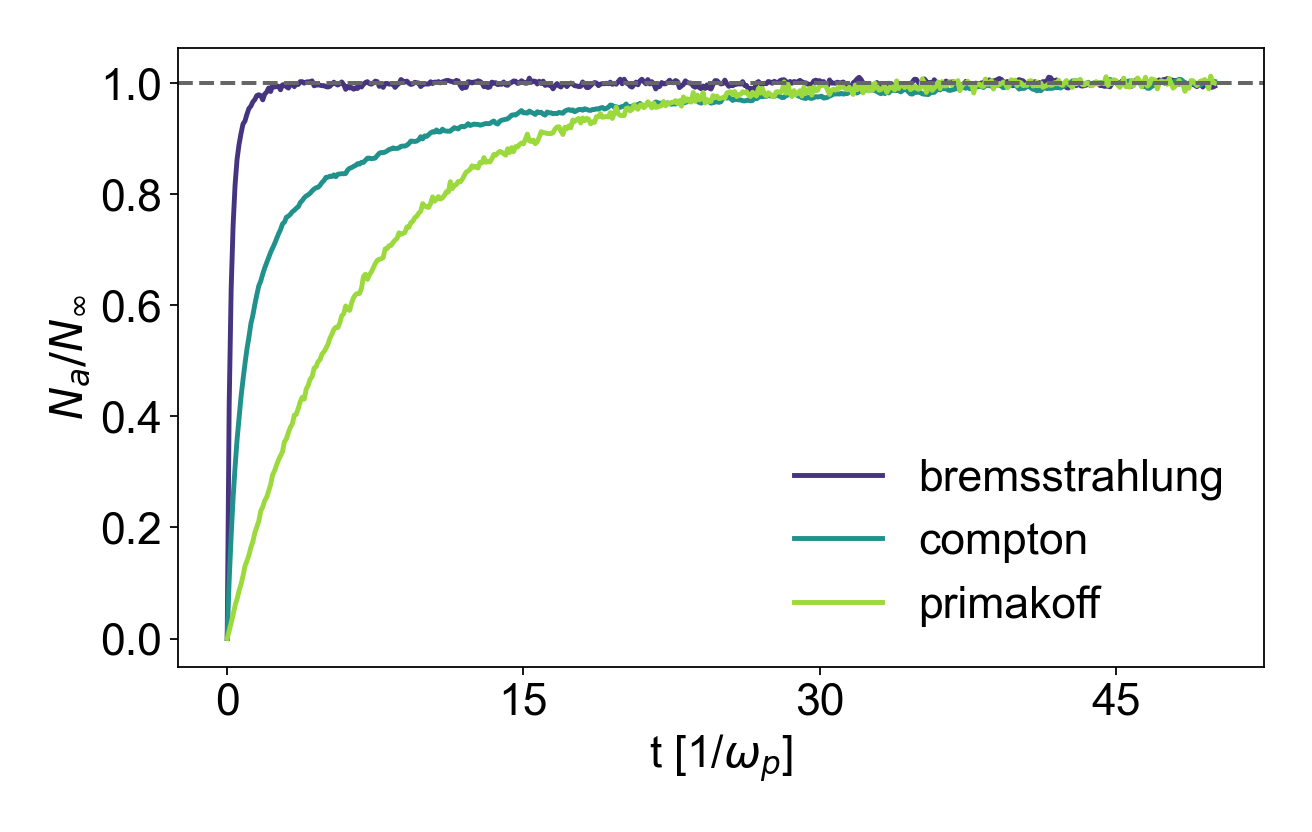}
    \caption{Relaxation of the weighted axion population toward equilibrium for bremsstrahlung, Compton-like production, and Primakoff conversion, shown as the normalised ratio $N_a/N_\infty$. Here $N_a(t)$ is the total axion population reconstructed from the dumps using macroparticle weights, and $N_\infty$ is the corresponding late-time steady-state value from the same run. The approach of all three curves toward unity demonstrates that the forward and inverse operators recover a stable population equilibrium.}
    \label{fig:eq_pop}
\end{figure}

\begin{figure}[t]
    \centering
    \includegraphics[width=\linewidth]{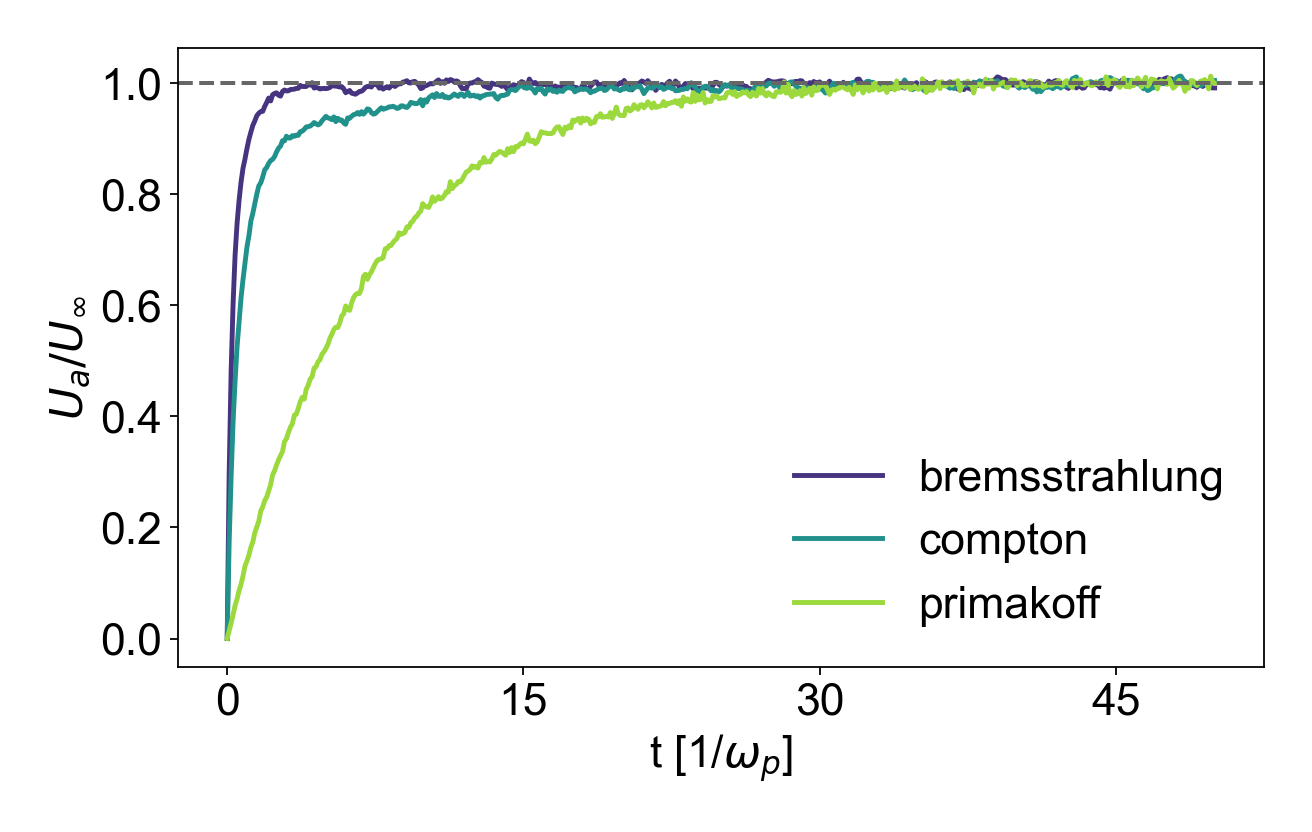}
    \caption{Relaxation of the weighted total axion energy toward equilibrium for bremsstrahlung, Compton-like production, and Primakoff conversion, shown as the normalised ratio $U_a/U_\infty$. Here $U_a(t)$ is the total axion energy reconstructed from the dumps using macroparticle weights and particle energies, and $U_\infty$ is the corresponding late-time steady-state value from the same run. The convergence of all three curves toward unity shows that the inverse operators also recover the correct steady-state energy balance.}
    \label{fig:eq_energy}
\end{figure}

To provide an initial validation of the inverse operators, we performed homogeneous thermal-cell relaxation tests for the three implemented channels separately: bremsstrahlung, Compton-like production, and Primakoff conversion. In each case, the corresponding forward and inverse processes were enabled simultaneously and the system was evolved from an initially axion free state toward equilibrium. Figures~\ref{fig:eq_pop} and \ref{fig:eq_energy} show the resulting approach to steady state in terms of the axion population and total axion energy, respectively.

The plotted quantities are reconstructed directly from the OSIRIS axion dumps using the macroparticle diagnostic weights. For each process, we compute the total axion population $N_a(t)$ as the weighted sum of axion macroparticle weights and the total axion energy $U_a(t)$ as the corresponding weighted sum of axion energies. We then normalise these by their late-time steady-state values, $N_\infty$ and $U_\infty$, obtained numerically from the late-time plateau of each run, so that the plotted ratios $N_a/N_\infty$ and $U_a/U_\infty$ approach unity at equilibrium. In this way, the figures test relaxation toward the correct steady state independently of the absolute normalisation of each individual process.

Reaching a steady state in both population and energy is an important validation of the inverse-process implementation. All three channels relax toward stable plateaus in both $N_a/N_\infty$ and $U_a/U_\infty$, demonstrating that the implemented inverse operators balance the corresponding forward source terms and recover the expected detailed-balance behavior in a homogeneous thermal environment. The different relaxation times simply reflect the different net equilibration rates of the three processes.

\section{Conclusion}

We have implemented a macroparticle-based axion extension within the OSIRIS QED--MC framework, including (i) an explicit axion species for transport and diagnostics, (ii) three axion-production channels widely used in thermal-plasma axion phenomenology---screened Primakoff conversion, Compton-like photoproduction, and electron bremsstrahlung with optional electron--electron contribution---and (iii) inverse operators, variance control, and optional conservative plasma-feedback capabilities. Benchmarking in homogeneous plasmas at $T_e=1.3$, 3, and 5~keV shows that the forward spectral emissivities reproduced by OSIRIS agree well with screened Raffelt-style analytic reference models, with percent-level agreement in integrated power and good agreement in spectral peak locations. We have also presented initial equilibrium-recovery tests showing that, for all three implemented channels, the coupled forward and inverse operators drive the axion population and axion energy toward stable steady-state values in a homogeneous thermal plasma.

These results establish the module as a practical kinetic framework for simulating axion production, absorption, and transport in homogeneous plasma environments within a mature PIC--MC code. In particular, the observed recovery of steady-state axion population and energy for all three implemented channels provides an initial validation that the inverse operators recover the expected detailed-balance behaviour in a thermal environment. At the same time, more extensive validation of conservative energy--momentum deposition, temperature evolution, and coupled multidimensional plasma scenarios remains important future work. 

\section*{Authors' contributions}
This project was conceived by GG and RB. MR has led the technical implementation of the OSIRIS axion--QED particle module. Both MR and AA have contributed to translating the axion physics into the OSIRIS code, with support from BM, PB, TG and LOS. The paper was written by MR with contributions from AA, PB, RB and GG.

\section*{Acknowledgements}
This research received funding from the UK Engineering and Physical Sciences Research Council (Grants No. EP/X01133X/1 and No. EP/X010791/1). AA and GG belong to the “Quantum Sensors for the Hidden Sector” consortium funded by the UK Science \& Technology Facilities Council (Grant No. ST/T006277/1). A.A. acknowledges the Saudi Arabian Cultural Bureau in the UK and the Rhodes Trust for funding his research and studies. The work of LOS was partially supported by FCT (Portugal) (grant X-MASER- 2022.02230.PTDC).

\appendix

\section{Bremsstrahlung convergence with axion macroparticle weight}

As an explicit validation of the weighted-emission and variance-control framework described in Sec.~II.D, we repeated the bremsstrahlung benchmark at $T_e=1.3$~keV using three different axion macroparticle weights, $w_a = 10^{-34}$, $10^{-36}$, and $10^{-38}$ in the OSIRIS macro-weight convention. All physical plasma parameters, couplings, and numerical settings were held fixed; only the coarse-graining of the emitted axion population into computational macroparticles was changed. The purpose of this test is to verify that the reconstructed physical emissivity is insensitive to the user choice of axion macroparticle weight.

In the formulation of Sec.~II.D.1, the expected number of \emph{physical} axions produced by an emitting electron macroparticle over one timestep is $\langle N_a^{\mathrm{phys}}\rangle = w_e \Gamma \Delta t_{\mathrm{phys}}$, where $w_e$ is the emitting-macroparticle weight, $\Gamma$ is the physical per-emitter axion-production rate, and $\Delta t_{\mathrm{phys}}$ is the physical timestep. This quantity does not depend on the chosen axion macroparticle weight $w_a$. Changing $w_a$ changes only how that same physical yield is represented computationally: smaller $w_a$ produces more simulated axion macroparticles of lower individual weight, whereas larger $w_a$ produces fewer, heavier macroparticles.

Equivalently, the expected number of axion macroparticle creation events is
\begin{equation}
\lambda_{\mathrm{macro}} = \frac{w_e \Gamma \Delta t_{\mathrm{phys}}}{w_a},
\end{equation}
so varying $w_a$ changes the expected macro-event count but not the underlying physical source term. When Poisson-mean capping is enabled, the sampled mean is replaced by $\lambda_{\mathrm{eff}} = \min(\lambda_{\mathrm{macro}},\lambda_{\mathrm{cap}})$ and the created-particle weight is rescaled according to $w_a' = w_a \lambda_{\mathrm{macro}}/\lambda_{\mathrm{eff}}$, as described in Sec.~II.D.2. This guarantees that the expected represented physical yield, $\mathbb{E}[K]\,w_a'$, remains equal to $w_e\Gamma\Delta t_{\mathrm{phys}}$. The convergence test therefore checks directly that changing the numerical coarse-graining does not change the reconstructed physics.

Figure~7 shows that the reconstructed bremsstrahlung spectral emissivity $dQ/dE$ is invariant, within the expected Monte Carlo noise, as $w_a$ is varied across four orders of magnitude, and that all three curves remain consistent with the analytic reference spectrum. As expected, the heaviest macroparticles produce the largest visible sampling noise, since the emitted population is then represented by fewer Monte Carlo particles, but no systematic shift in spectral shape or normalization is observed.

The PIC spectrum is reconstructed by incrementally accumulating the energy histogram of newly appended axions across successive dumps in order to avoid double counting. Diagnostic macroparticle weights are converted to physical particle number using the reference density and cell volume, and the resulting weighted energy histogram is normalized by the 3D simulation volume and total physical runtime to obtain $dQ/dE$ in cgs units.

\begin{figure}[H]
\centering
\includegraphics[width=\linewidth]{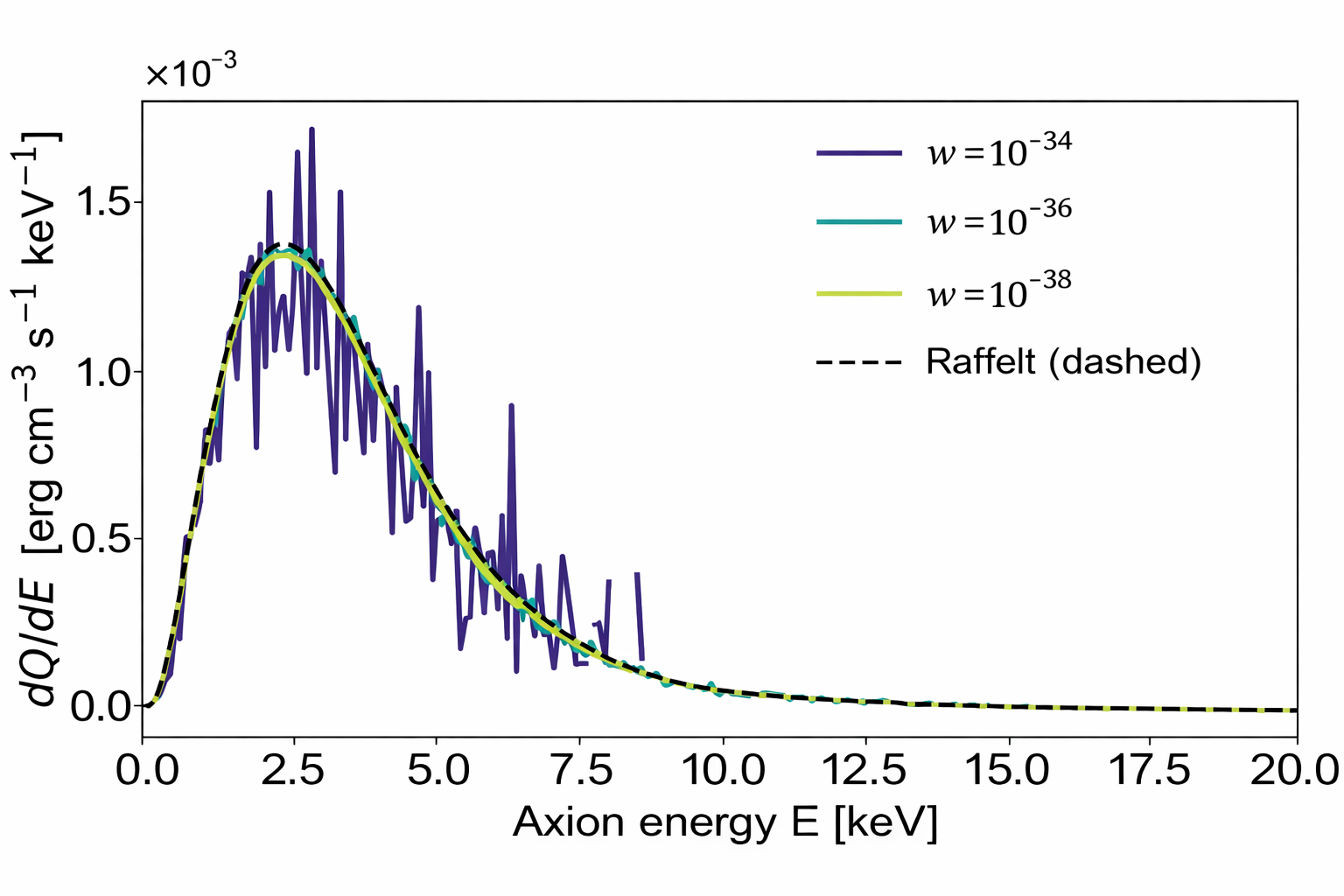}
\caption{Bremsstrahlung spectral-emissivity convergence test at $T_e=1.3$~keV for three axion macroparticle weights, $w_a = 10^{-34}$, $10^{-36}$, and $10^{-38}$. The overlap of the reconstructed PIC spectra demonstrates that the emitted spectrum is invariant, up to Monte Carlo noise, under changes in axion macroparticle coarse-graining, as expected for an unbiased weighted source operator. The dashed curve is the analytic reference spectrum used in Sec.~III.}
\label{fig:brems_convergence_waxion}
\end{figure}

\section{Reconstructing spectra from OSIRIS macroparticles}
The PIC spectra shown below are constructed from OSIRIS particle dumps.
Newly created axion macroparticles between consecutive dumps are identified using a length-difference rule: if the axion container size grows from $N_{j-1}$ to $N_j$, the indices $i\in[N_{j-1},N_j)$ are treated as newly emitted.
Treating axions as massless, the axion energy is computed from the OSIRIS-normalized momentum components as
\begin{equation}
E_{\mathrm{keV}} = (m_ec^2)_{\mathrm{keV}}\,|\bm{p}| = 511\,\sqrt{p_1^2+p_2^2+p_3^2}.
\label{eq:Ek_keV}
\end{equation}
To construct a \emph{power} spectrum we weight each emitted axion by its carried energy, $w_i=|q_i|E_i$, where $q_i$ is the diagnostic macroparticle weight recorded in the OSIRIS particle dump, i.e.\ the number of physical axions represented by macroparticle $i$, and histogram in energy bins,
\begin{equation}
H_k = \sum_{i\in \mathrm{bin}\,k} |q_i|\,E_i.
\end{equation}
For shape comparisons we plot the normalised spectrum
\begin{equation}
\left.\frac{1}{Q}\frac{\dd Q}{\dd E}\right|_{E_k}
\approx
\frac{H_k}{\sum_j H_j\,\Delta E_j},
\end{equation}

Here $Q$ denotes the total volumetric emissivity over the plotted energy range,
\begin{equation}
Q \equiv \int \frac{dQ}{dE}\, dE,
\end{equation}
so that $Q^{-1} dQ/dE$ is the normalised spectral emissivity. Then, $\sum_k (Q^{-1}\dd Q/\dd E)_k\,\Delta E_k=1$.
For absolute emissivities we divide by the total physical run time and emitting volume,
\begin{equation}
\left.\frac{\dd Q}{\dd E}\right|_{E_k}
\approx
\frac{H_k}{V\,t_{\mathrm{tot}}\,\Delta E_k}\times (1\,\mathrm{keV}),
\label{eq:absolute_spectrum}
\end{equation}
using $1\,\mathrm{keV}=1.602\times10^{-9}\,\mathrm{erg}$ to report $\dd Q/\dd E$ in cgs units.
Conversion between OSIRIS normalised units and physical volume/time follows standard OSIRIS conventions \cite{fonseca_2002_osiris}

\end{document}